\documentclass{WileyMSP-template}
\usepackage{hyperref}
\begin{document}

\title{Autonomous X-ray Fluorescence Mapping of Chemically Heterogeneous Systems via a Correlative Feature Detection Framework}

\maketitle


\author{Carlos Deleon}
\author{Dmitri Gavrilov}
\author{Peggy O'Day}
\author{Ajith Pattammattel*}



\begin{affiliations}
Carlos Deleon \\
Department of Computer Science, Stony Brook University, Stony Brook, NY, USA \\

Dr. Ajith Pattammattel, Dr. Dmitri Gavrilov \\
National Synchrotron Light Source II, Brookhaven National Laboratory, Upton, NY, USA \\
pattammattel@bnl.gov

Dr. Peggy O'Day \\
Life and Environmental Sciences Department and the Sierra Nevada Research Institute, University of California, Merced, California 95343, United States
Environmental Systems Graduate Program, University of California, Merced, 95343, United States \\
pattammattel@bnl.gov

\end{affiliations}


\keywords{Autonomous Scanning, Synchrotron Imaging, Beamtime Optimization, Computer Vision}

\begin{abstract}

We present X-AutoMap, a modular framework for autonomous X-ray fluorescence (XRF) mapping that enables chemically informed targeting of regions of interest through a correlative feature detection strategy. The system integrates classical computer vision and rule-based logic to identify features based on spatial relationships across multiple elemental maps, rather than relying solely on intensity or morphology. Tight integration with the Bluesky control infrastructure at the NSLS-II Hard X-ray Nanoprobe (HXN) beamline enables real-time, closed-loop scan orchestration. Applied to a chemically heterogeneous urban PM$_{2.5}$ sample, X-AutoMap reduced high-resolution acquisition time from over 44 hours to approximately 10 hours by targeting compositionally significant features identified from coarse scans. High-resolution results revealed diverse particle types—including fully mixed, partially overlapping, and spatially distinct multi-element structures—demonstrating the system’s ability to isolate chemically relevant features with minimal user intervention. The framework supports interactive and autonomous modes, operates within hardware constraints via grid-based scanning, and is robust across varying sample conditions. Future extensions will incorporate machine learning and probabilistic sampling to further improve detection sensitivity and scan efficiency. X-AutoMap is currently in active use at HXN and provides a flexible foundation for scalable, intelligent imaging workflows at synchrotron beamlines.

\end{abstract}


\section{Introduction}

Understanding the spatial distribution of elements and their chemical states is fundamental across a wide range of disciplines, including geochemistry, environmental science, materials engineering, and planetary research. Imaging plays a central role in these fields, enabling direct observation of how chemical and structural features vary across complex systems. In both natural and synthetic materials, localized features—such as grain boundaries, nanoscale inclusions, and chemically distinct domains—often govern reactivity, transport, and overall system behavior. These features influence processes ranging from contaminant mobility to catalytic efficiency. To develop a comprehensive understanding, researchers often examine large number of samples or survey extended areas to capture meaningful variability and statistically significant trends. This requires imaging techniques that are not only chemically specific and spatially resolved, but also scalable and efficient enough to support high-throughput or large-area analysis.

Synchrotron-based X-ray microscopy has emerged as a powerful technique for probing such systems, combining high spatial resolution, deep penetration, and sensitivity to both elemental composition and chemical state.\cite{sakdinawat_nanoscale_2010} Recent advances in nanoprobe instrumentation have pushed spatial resolution to the ten-nanometer scale and detection limits to hundreds of atoms.\cite{masteghin_benchmarking_2024} These capabilities have opened new opportunities in materials characterization, enabling the visualization of chemical and elemental heterogeneity at previously inaccessible scales.\cite{pattammattel_hard_2021,michelson_three-dimensional_2022} However, the serial nature of data acquisition—where each pixel is measured individually—poses practical limitations. For example, scanning a ten by ten micrometer region at ten-nanometer resolution requires one million measurements and several hours of instrument time.

This bottleneck forces users to compromise between resolution, scan area, and acquisition time. Most workflows rely on low-resolution survey scans to manually identify regions of interest for higher-resolution imaging. This process is time-consuming, subjective, and increasingly misaligned with the complexity of modern samples and the constraints of shared beamline environments. These challenges have motivated the development of autonomous workflows that can intelligently guide data collection based on real-time feedback.

Automated scanning strategies aim to mimic expert behavior: surveying broadly, identifying candidate features, and performing focused follow-up scans. By integrating live image analysis with instrument control, such systems can dynamically prioritize regions that are likely to be informative. This approach is particularly valuable for heterogeneous samples, such as environmental particulates, mineral aggregates, or engineered materials, where critical features may be chemically distinct but spatially sparse or irregularly distributed.\cite{pattammattel_iron_2021} Automating their detection improves throughput, enhances reproducibility, and reduces user bias.\cite{li2025advancing}

While many existing approaches rely on spatial features—such as shape, intensity, or texture—techniques like XRF mapping introduce additional complexity.\cite{germinario2016textural,chowdhury2022roi} Each pixel in an XRF scan contains a full energy-dispersive spectrum, which is fitted to extract elemental composition. This hyperspectral structure means that relationships across the spectral (photon energy) dimension are often as important as spatial features. In practice, features of interest are frequently defined not by intensity alone, but by the correlation, co-localization, or mutual exclusion of specific elements—signatures that reflect redox gradients, mineralogical boundaries, or anthropogenic contamination. These patterns may not be visually apparent and are easily missed by detection methods that operate on single-element maps or rely solely on image-based cues.

This highlights a broader challenge for autonomous imaging systems: how to combine spectral reasoning across multiple elemental channels with spatial detection in a way that is fast, interpretable, and adaptable. Such integration is particularly important at scientific user facilities, where experimental needs vary and beamtime is limited. Recent advances in modular control systems, real-time data processing, and algorithmic decision-making now make it possible to construct flexible, closed-loop scanning pipelines that respond to evolving measurements during an experiment.

In this work, we present \textit{X-AutoMap}, a modular and configurable framework for autonomous scan orchestration in XRF microscopy, centered around a correlative feature detection strategy. The system uses classical computer vision techniques\cite{bradski2008learning} to identify features independently from multiple elemental maps and applies logical operations to select regions based on user-defined criteria for spatial correlation or anti-correlation (Figure~\ref{fig:workflow}). This mirrors how expert users interpret elemental distributions to target regions for further analysis, but formalizes that intuition into a reproducible, automated process. We demonstrate the approach using a chemically heterogeneous PM$_{2.5}$ sample analyzed at the NSLS-II Hard X-ray Nanoprobe beamline.\cite{nazaretski_design_2017} X-AutoMap enables real-time targeting of chemically meaningful regions, supports both interactive and fully autonomous modes, and forms a generalizable foundation for intelligent data collection in hyperspectral imaging workflows.

\section{Results and Discussion}
\subsection*{Correlative Feature Detection}

\begin{figure}
  \centering
  \includegraphics[width=\linewidth]{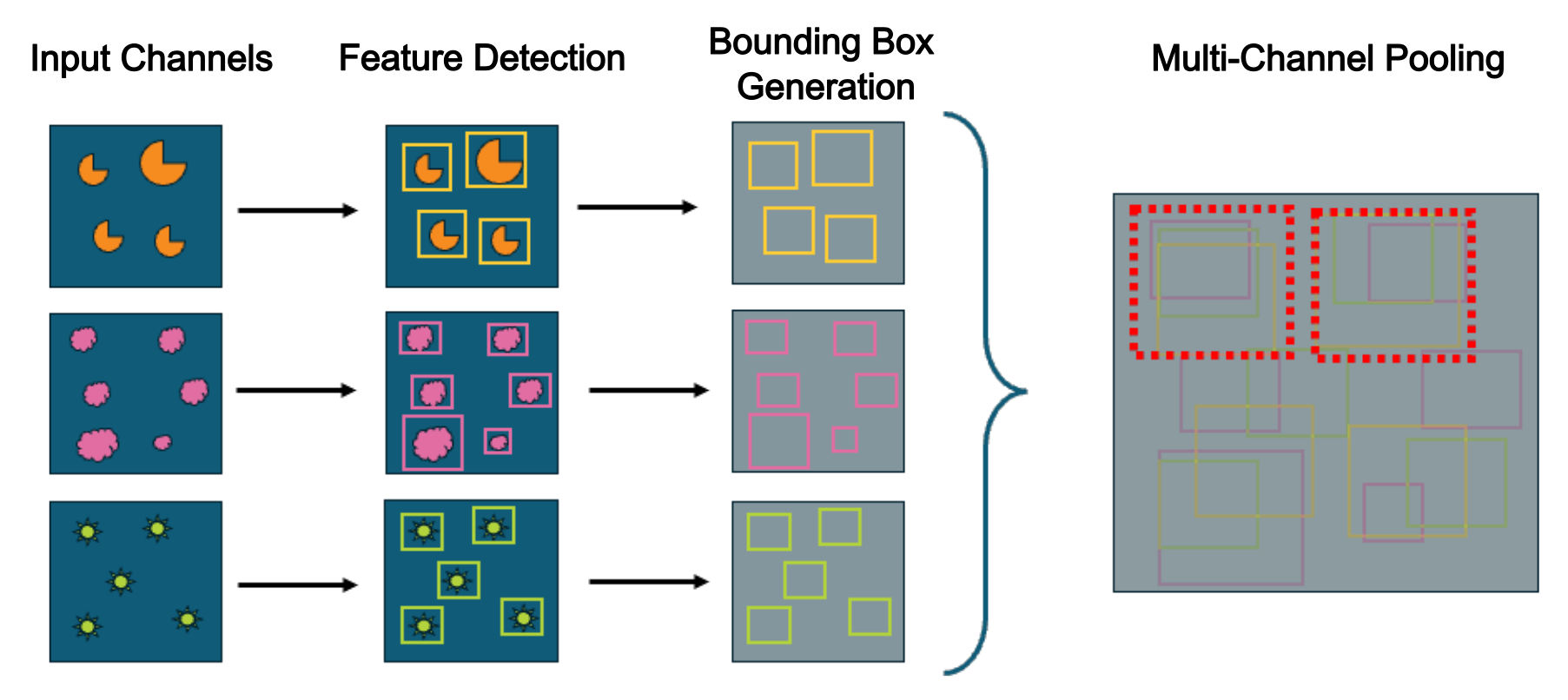}
  \caption{Multi-modal feature detection and ROI merging strategy used for autonomous targeting. Features are first detected independently from multiple input channels (e.g., elemental maps) using classical computer vision methods. Each detected feature is converted into a standardized bounding box based on the scan resolution. Bounding boxes from all channels are then spatially merged and clustered to identify regions of multimodal significance. The final set of high-priority scan targets (highlighted in red) reflects spatial overlap or correlation between features across different channels.}
  \label{fig:workflow}
\end{figure}

To enable autonomous targeting of chemically relevant regions, we developed a correlative feature detection strategy that identifies regions of interest (ROIs) based on spatial relationships across multiple elemental distributions in coarse-resolution XRF maps (Figure~\ref{fig:workflow}). Unlike traditional methods that focus on visual features such as shape or intensity, this approach prioritizes chemically meaningful relationships—specifically, how different elemental species co-locate or exclude one another. By capturing these correlations, the strategy reflects underlying geochemical or environmental processes that are not readily apparent through intensity-based detection alone.

Each elemental map is processed independently using classical computer vision methods, including adaptive thresholding and blob detection. Morphological dilation is applied to enhance faint or fragmented signals, improving the detectability of low-concentration features. Detected ROIs are converted into standardized bounding boxes to maintain compatibility with scan control protocols and beamline resolution constraints.

Logical operations are applied across the spatially aligned bounding boxes to evaluate compound relationships such as co-localization (e.g., Fe \texttt{AND} Cr) or exclusion (e.g., Cr \texttt{NOT} Si). Clustering is used to merge overlapping ROIs across channels, resulting in a consolidated set of prioritized scan targets. This interpretable, rule-based approach allows researchers to encode domain knowledge directly into the detection logic, enabling reproducible and scientifically informed decision-making without requiring prior training data or machine learning models.

\subsection*{Pre-Experiment Optimization and Interface Design}

To ensure robust performance at the beamline, we conducted offline testing using archived XRF datasets to tune key parameters in the OpenCV pipeline. Among several detection algorithms evaluated, the SimpleBlobDetector was selected for its robustness to noise and configurability across different signal conditions. Morphological preprocessing and dynamic thresholding were critical in adapting the detection pipeline to the sparse and heterogeneous nature of the PM$_{2.5}$ data.

A graphical user interface (GUI) was developed to support interactive parameter tuning and real-time ROI preview. This tool enables users to explore combinations of intensity and area thresholds, visualize ROI overlays on elemental maps, and evaluate the impact of different logic rules before deployment. The GUI was used both during development and in situ at the beamline to adapt detection parameters to live scan conditions. The full toolchain, including detection, logic, and orchestration modules, is available through the open-source X-AutoMap repository on GitHub

\subsection*{Integrated Beamline Orchestration}

To realize closed-loop scanning, the correlative detection system was integrated with the NSLS-II HXN beamline using the Bluesky control framework. Upon completion of an initial coarse scan, a monitored EPICS process variable triggered the feature detection pipeline. Detected ROIs were formatted as JSON scan plans and automatically submitted to the Bluesky Queue Server (QServer). 

All follow-up scans were initiated by the beamline without user intervention. Scan positions and motor limits were computed directly from the bounding box coordinates of the detected ROIs. The full cycle—from coarse scan completion to fine scan execution—typically required less than one minute. This orchestration capability enabled extended autonomous operation, including overnight acquisition sessions, with minimal oversight.

The system also supports hybrid workflows, where users can interactively refine parameters mid-session and requeue updated scan plans. This flexibility was particularly useful when transitioning from one region of the sample to another with different signal characteristics.

\begin{figure}
  \includegraphics[width=\linewidth]{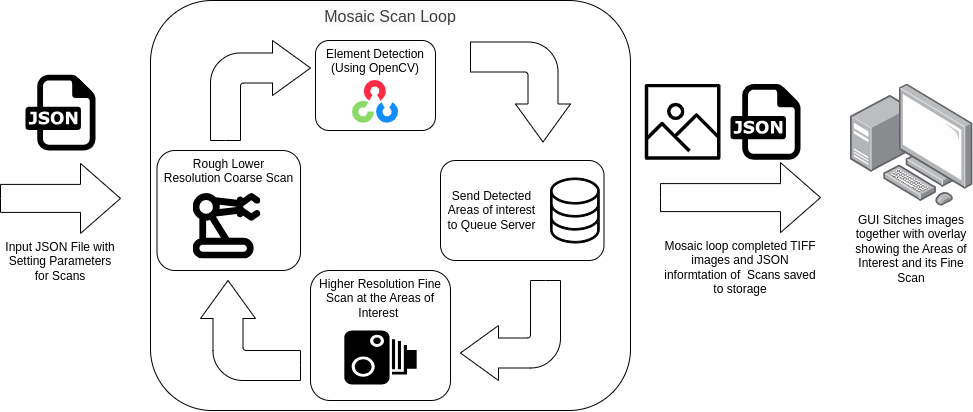}
  \caption{Overview of the X-AutoMap autonomous scanning workflow. The system operates in a loop beginning with a coarse, low-resolution XRF scan of a mosaic region. Elemental features are detected in real time using OpenCV-based computer vision, and the resulting regions of interest (ROIs) are passed to a queue server for targeted high-resolution scanning. All scans and detected ROIs are saved as JSON files. A graphical user interface (GUI) then stitches the coarse and fine scans, overlaying the ROIs for visualization and verification. This closed-loop approach enables automated, real-time targeting of chemically relevant features in large-area scans.}
  \label{fig:boat1}
\end{figure}

\subsection*{Application to PM$_{2.5}$ Sample Analysis}

We evaluated the complete system using a chemically complex urban PM$_{2.5}$ sample, chosen for its diverse and spatially heterogeneous particle composition. Elements such as Fe, Si, Cu, Cr, and Ca were variably distributed across submicron particulates. A coarse mosaic scan ($200 \times 200~\mu$m$^2$, 500 nm step, 5 ms dwell) was first acquired to locate regions with compositional variability.

Using correlative logic across Fe, Ca, and Si maps, the detection algorithm selected ROIs based on user-defined criteria for co-localization and exclusion. For example, regions containing Fe and Ca but lacking Si were prioritized, reflecting domains of likely anthropogenic origin. These ROIs were automatically scanned at high resolution (50 nm step, 10 ms dwell), revealing submicron features with distinct chemical environments.

Compared to a uniform high-resolution scan over the same area, which would require over 44 hours of beamtime, the correlative targeting strategy reduced total scan time to approximately 10 hours. While approximately 20\% of potential features were missed—primarily due to low signal strength or logical rule boundaries—the trade-off was considered acceptable given the throughput gains.

\begin{figure}
  \centering
  \includegraphics[width=0.75\linewidth]{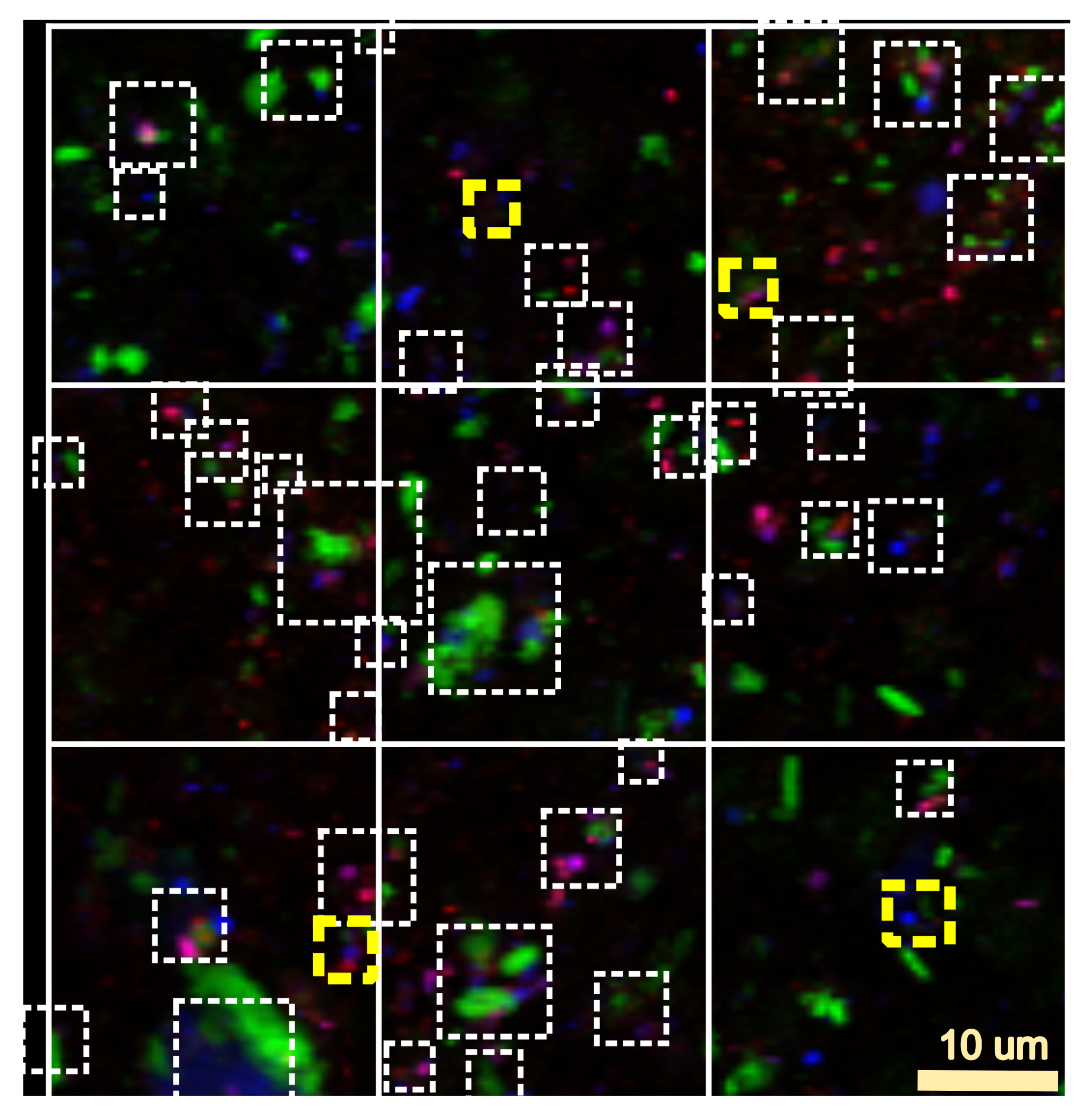}
  \caption{Example of autonomous region selection using X-AutoMap. Composite XRF map of a chemically heterogeneous PM$_{2.5}$ sample, with Fe (red), Ca (green), and Si (blue) elemental distributions. White dashed boxes indicate regions automatically selected for high-resolution scanning based on correlative feature detection. Yellow dashed boxes highlight representative features that were not selected by the autonomous workflow. This image illustrates both the effectiveness and current limitations of the correlative detection strategy in identifying chemically relevant features across large, complex scan areas.}
  \label{fig:finescans}
\end{figure}

\subsection*{Performance and Observations}

High-resolution XRF scans revealed distinct types of chemical and spatial patterns among the detected particles. These could be broadly grouped into three categories based on the degree of elemental mixing: (1) adjacent but unmixed features, (2) fully co-localized multi-element structures, and (3) partially overlapping distributions (Figure~\ref{fig:finescans}). Type 1 particles displayed physically aggregated yet chemically distinct phases, such as Ca- and Si-rich domains in proximity to Fe hotspots. Type 2 particles exhibited strong co-localization of all three elements, suggesting chemically mixed or internally heterogeneous particles. Type 3 particles showed partial overlap between two elements with the third spatially segregated, indicating complex formation or partial surface mixing.

These examples demonstrate the effectiveness of the correlative detection logic in isolating chemically diverse targets across a heterogeneous landscape. In many cases, features that were difficult to interpret in the coarse-resolution maps—due to low contrast or background interference—were revealed in much greater detail in the follow-up scans. Moreover, the ability to encode domain knowledge through logical rules allowed users to suppress common interferences, such as Si-dominated substrate regions, while still capturing relevant particle features.

The scan orchestration pipeline remained stable and responsive throughout multi-session testing. ROI selection proved robust to a range of parameter settings, and the graphical interface enabled intuitive tuning and preview during both pre-beamtime testing and live acquisition. The system’s support for both interactive and autonomous operation modes allowed users to adapt to varying levels of sample complexity and prior knowledge.

Together, these results validate the X-AutoMap framework as a flexible and efficient tool for chemically informed scanning. It demonstrates the successful integration of correlative feature detection with real-time beamline orchestration and provides a foundation for further extensions, including adaptive prioritization and machine learning–driven targeting.

\section{Methods}

\subsection*{Experimental Setup}

All measurements were performed at the Hard X-ray Nanoprobe (HXN) beamline at the National Synchrotron Light Source II (NSLS-II), Brookhaven National Laboratory. For this study, XRF scans were performed at 12 keV using a four-element silicon drift detector (SDD) positioned at 90 degrees to the incident beam. Unfitted XRF ROI maps were used for ROI selection in this paper based on known elemental emission lines. Urban PM$_{2.5}$ samples were collected on teflon filters and then mounted on silicon windows for X-ray imaging.

\subsection*{Autonomous Coarse-to-Fine Scanning Workflow}

X-AutoMap implements a modular, coarse-to-fine scanning workflow integrated with the control infrastructure at the NSLS-II beamlines, the Bluesky user interface (BSUI), and the Bluesky Queue Server (\texttt{QServer})\cite{rakitin2022next}. The workflow consists of two main stages: (1) an initial coarse-resolution XRF scan to survey the sample, and (2) automated high-resolution scans at selected regions of interest (ROIs). The system is built around JSON-based plan (example shown in the supporting information) definitions and real-time interaction with \texttt{QServer}. The complete source code is available on GitHub{\url{https://github.com/CodingCarlos23/X-AutoMap.git}}.

Coarse scans are submitted to \texttt{QServer} from a BSUI IPython kernel, typically using a step size of 500~nm and a dwell time of 5~ms per point (both user-defined). Due to piezo motor limitations at the HXN beamline, the maximum scan range for a single acquisition is constrained to $25 \times 25~\mu$m$^2$. To enable coverage of larger regions, X-AutoMap implements a grid scanning strategy in which adjacent tiles are acquired sequentially to construct a full mosaic. Typical total scan areas range from $100 \times 100~\mu$m$^2$ to $200 \times 200~\mu$m$^2$, depending on the experimental context and region of interest.

Upon completion of each tile scan, elemental maps are generated based on user-defined spectral regions of interest and saved as \texttt{.tiff} images. These serve as input to the correlative feature detection pipeline, which processes each elemental channel independently using OpenCV. Morphological dilation is first applied to enhance weak or disconnected signals, followed by blob detection using \texttt{cv2.SimpleBlobDetector}. ROIs are then filtered and merged using logical operations defined by the user (e.g., Fe \texttt{AND} Cr \texttt{NOT} Si) to identify chemically relevant targets.

Detected ROIs are converted into standardized bounding boxes and compiled into a JSON document that specifies motor positions, scan dimensions, detector configurations, and metadata required for high-resolution follow-up. These scan plans are immediately submitted to QServer, where they are executed as fly scans with a typical step size of 50~nm and a dwell time of 10~ms per point (User defined). All scan parameters, execution metadata, and results are saved to the user data directory for downstream analysis.

Once all high-resolution scans for a given tile are complete, the system advances to the next tile in the mosaic and repeats the process. This tile-by-tile coarse-to-fine loop continues until a user-defined stopping condition is met, such as a target area (used in this work), total scan time, or number of ROIs acquired.

This architecture enables fully autonomous, real-time operation within the constraints of the beamline’s motion system. The complete cycle—including coarse scan, feature detection, plan generation, and fine scan execution—typically completes in under 10 minute per tile (depending of number of ROIs found). This allows the system to operate efficiently during extended or unattended sessions, supporting high-throughput imaging across large sample areas with minimal human oversight.

\begin{figure}
  \centering
  \includegraphics[width=0.75\linewidth]{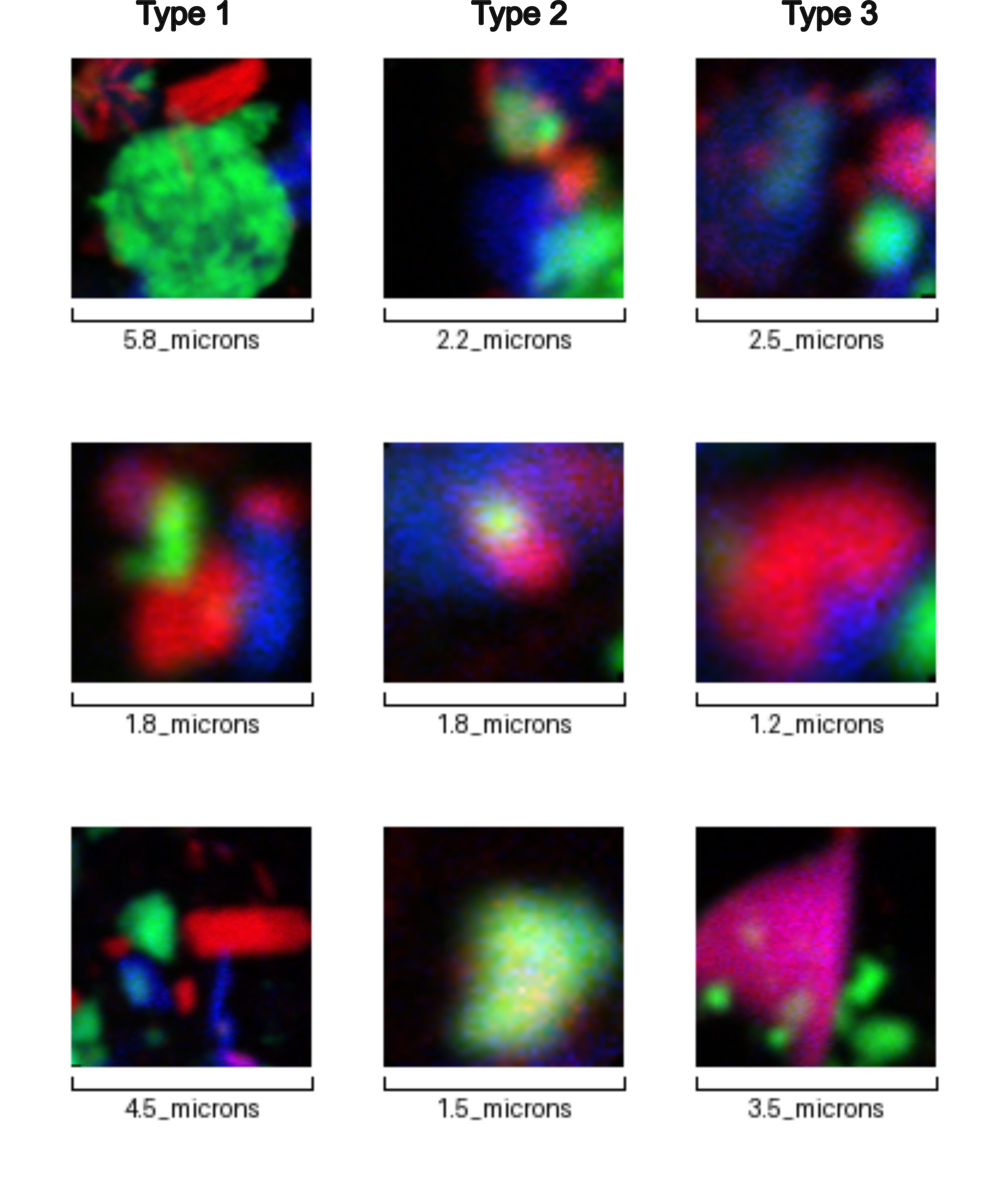}
  \caption{Examples of high-resolution XRF scans autonomously selected by X-AutoMap, illustrating different types of elemental correlations and mixing behaviors in PM$_{2.5}$ particles. Each column represents a distinct particle class based on spatial and chemical relationships among Fe (red), Ca (green), and Si (blue). Type 1 shows adjacent but largely unmixed elemental regions, indicating physical aggregation of chemically separate particles. Type 2 particles show strong co-localization of all three elements, suggesting chemically or physically mixed phases. Type 3 features exhibit partial overlap between two elements with the third spatially distinct.  Scale bars indicate approximate particle dimensions in microns. These variations highlight the diversity of particle compositions and the effectiveness of the correlative feature detection strategy in targeting chemically informative structures.}
  \label{fig:finescans}
\end{figure}
\subsection*{Post-Processing and Visualization}

Following acquisition, all elemental maps and scan metadata were processed using a custom GUI developed for X-AutoMap. The interface allows users to visualize stitched coarse and fine maps, overlay detected ROIs, and evaluate targeting performance. The GUI also enables real-time adjustment of blob detection parameters and thresholding values, supporting both interactive and automated use modes. The complete software stack, including GUI, detection pipeline, and queue handler, is available at \url{https://github.com/CodingCarlos23/X-AutoMap.git}.



\section{Conclusion}
In this work, we presented \textit{X-AutoMap}, a modular and extensible framework for autonomous X-ray fluorescence (XRF) mapping, centered around a correlative feature detection strategy. The system integrates rule-based logic with classical computer vision to identify chemically relevant regions based on spatial relationships across multiple elemental maps. Through tight integration with the Bluesky control framework and real-time scan orchestration, X-AutoMap enables closed-loop operation directly at the beamline.

We validated the system on a chemically heterogeneous urban PM$_{2.5}$ sample, chosen for its compositional complexity and environmental relevance. Using correlative logic across Fe, Ca, and Si maps, X-AutoMap successfully identified spatially distinct or overlapping chemical domains in submicron particles. The system was able to differentiate between co-localized, partially mixed, and physically adjacent but chemically distinct particle types, offering detailed insight into the compositional diversity within a single field of view. These particle-level features—difficult to identify manually or through single-channel logic—were rapidly and reproducibly targeted by the autonomous workflow.

Compared to a full-area high-resolution scan, which would require more than 44 hours of beamtime, our correlative targeting strategy reduced total acquisition time to approximately 10 hours—a more than fourfold improvement. Although approximately 20\% of expected features were missed, often due to low-intensity signals or logic rule exclusions, the trade-off was acceptable given the substantial gains in throughput and automation. The framework also enabled suppression of background-dominated areas, such as the Si-rich filter substrate, allowing for selective acquisition of chemically meaningful structures.

Future development will focus on augmenting the current rule-based strategy with machine learning approaches capable of learning from spatial, spectral, and statistical correlations across large datasets. This includes improved noise tolerance, real-time threshold adjustment, and integration of Bayesian models to guide sampling decisions dynamically across tiles. Such extensions could enable probabilistic scan prioritization, increasing the likelihood of capturing rare or low-abundance particle types.

X-AutoMap is currently deployed at the NSLS-II Hard X-ray Nanoprobe beamline, where it supports autonomous scanning workflows across a variety of material systems. This work demonstrates that lightweight, interpretable automation strategies—when coupled with robust beamline integration and chemically aware targeting—can significantly enhance the scalability, precision, and intelligence of synchrotron-based imaging workflows. The modular approach X-AutoMap would allow future integration of advanced feature detection methods for autonomous scanning.\cite{harraden2025mineralogy}  in As the complexity of samples and experimental demands continue to grow, frameworks like X-AutoMap offer a critical path forward for enabling efficient, data-driven discovery at modern user facilities.




\medskip
\textbf{Acknowledgments} \par 
This research used HXN-3ID of the National Synchrotron Light Source II, a U.S. Department of Energy (DOE) Office of Science User Facility operated for the DOE Office of Science by Brookhaven National Laboratory under Contract No. DE-SC0012704. The authors thank Dr. Carlos Soto for fruitful discussions. 


%
\bibliographystyle{MSP}
\bibliography{references}

\end{document}